\documentclass[twocolumn,prl,aps,letterpaper,groupedaddress]{revtex4-1}

\usepackage{graphicx}
\usepackage{epstopdf}
\usepackage{bm}
\usepackage{amsmath,amssymb}
\usepackage{color}
\usepackage{caption}
\usepackage{array}

\newcommand{\chia}[1]{{\color{black}{#1}}} 
\newcommand{\mc}[1]{{\color{black}{#1}}}

\newcommand{\rev}[1]{{\color{black}{#1}}}

\begin{document}
\title{Demonstration of measurement-only blind quantum computing}
\author{Chiara Greganti$^1$, Marie-Christine Roehsner$^1$,  Stefanie Barz$^{1,2}$, Tomoyuki Morimae$^3$, Philip Walther$^1$}
\affiliation{$^1$~University of Vienna, Faculty of Physics, Austria, $^2$~Present address: University of Oxford,  Clarendon Laboratory, UK, $^3$~ASRLD Unit, Gunma University, 1-5-1 Tenjin-cho Kiryu-shi Gunma-ken, 376-0052, Japan}

\date{\today}
\begin{abstract}
Blind quantum computing allows for secure cloud networks of quasi-classical clients and a fully-fledged quantum server. Recently, a new protocol has been proposed, which requires a client to perform only measurements. We demonstrate a proof-of-principle implementation of this measurement-only blind quantum computing, exploiting a photonic set-up to generate four-qubit cluster states for computation and verification. \chia{Feasible}  technological requirements for the client and \chia{the device-independent blindness} make this scheme very applicable for future secure quantum networks.
\end{abstract}
\maketitle

\textit{Introduction.} -- Quantum physics enables \rev{one} to enhance security for processing data over a distributed network. 
In particular, quantum cloud computing allows quasi-classical clients \rev{(i.e. clients with a limited amount of quantum resources, such as qubit preparation or detection)} to  
\rev{ do calculations beyond their computational power, namely perform quantum algorithms.}
  The first proposed and demonstrated two-party secure quantum cloud computation protocol is known as blind quantum computing (BQC) \cite{BFK}: a client, Alice, who can generate only single-qubit states, delegates 
her quantum computing to a remote server, Bob, who has a fully-fledged quantum computer,
without leaking any of her privacy. 
 Many  theoretical studies based on this have been performed recently
~\cite{MABQC,FK,Vedran,AKLTblind,topoblind,CVblind,Lorenzo,Joe_intern,Sueki,tri}, and also experimental demonstrations have been reported \cite{Barz, BarzNP,Fisher14}.  
A  simplified and novel version for secure quantum computing consists of a two-party protocol \cite{MABQC,topoveri}, where   Alice only makes  measurements and  Bob's blindness is proven by the no-signaling principle \cite{PR}.  Here blindness indicates that whatever Bob does, he cannot learn any of Alice's privacy.

In order to underline the feasibility of the measurement-only BQC 
we demonstrate the computation protocol in a photonic experiment.  Bob generates four-qubit resource states that are used by Alice to implement generic two-qubit entangling gates and verification protocols.\\

\begin{figure}[b]
\begin{center}
\includegraphics[width=0.35\textwidth]{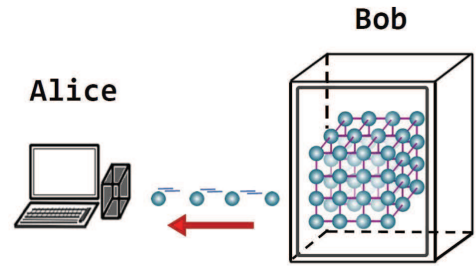}
\end{center} \captionsetup{justification=raggedright}
\caption{Measurement-only blind quantum computing.
Alice, a client with a classical device (laptop) and a quantum  detector device (grey box), receives quantum information via a one-way channel from Bob, a quantum server capable of generating entangled quantum states as universal resources. 
Here Bob's resource corresponds to a fault-tolerant 3D-cluster state, but it can be any other measurement-based quantum computation resource.
} 
\label{fig2}
\end{figure}

\textit{Theory.} --  
The idea of  
measurement-only BQC 
is shown in Fig.\ref{fig2}. 
Bob generates a resource state for measurement-based quantum computing (MBQC)~\cite{Raussendorf2001}\rev{~\cite{Briegel2009}},
and sends the corresponding qubits, one by one,  through a one-way quantum channel
 to Alice. 
She measures each qubit according to her program. 
For any kind of a malicious Bob, he cannot learn anything about Alice's quantum computation, because information \chia{is sent only} in one direction. The no-signaling principle then ensures that if Alice and Bob share a system (classical, quantum, or superquantum) and she measures her part, this does not transmit any information to Bob.
This principle is more fundamental than quantum physics  \cite{PR} and consequently provides security even against superquantum attacks. 
\begin{figure*}[t]
\begin{minipage}[c]{0.68\textwidth}
    \includegraphics[width=1\textwidth]{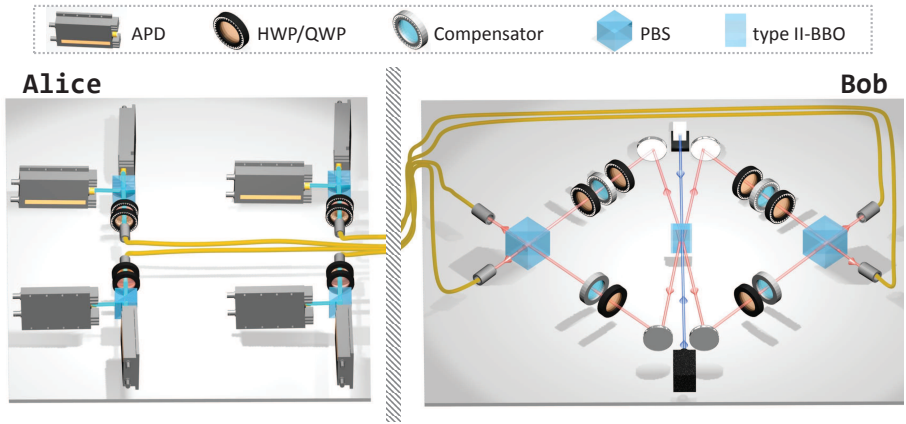}
  \end{minipage}\hfill
  \begin{minipage}[c]{0.29\textwidth} \captionsetup{justification=raggedright}
    \caption{
      Schematic Alice's (left) and Bob's (right) optical set-ups. Right: The four-qubit entangled states  are generated via a fs-pulsed laser pumping a type-II BBO crystal in a double-passage scheme. The compensator, HWPs and PBSs allow Bob to select the desired four-qubit resource. Left: The detection requires HWP, QWP and PBS for measuring in different polarization basis, and single-photon APD per qubit. The laboratories are connected by single-mode fibers as quantum channels. See main text for acronyms.
    } \label{setup}
  \end{minipage}
\end{figure*}

We  remark that the original BQC  examines a  different approach.  A quasi-classical Alice must be able to:  generate randomly-rotated single qubit states, send these via a quantum channel to  Bob, and interact via classical channels in order to control and receive the computation results.  \chia{Recently, the single-qubit generation requirement is extended to the coherent state generation~\cite{Vedran} or single-qubit measurement~\cite{Hajdusek}.  }
 The measurement-only BQC scheme centers on a quasi-classical Alice, who now only  receives qubits for measuring those. This alternative concept produces a 
 practical computing protocol. \chia{In particular in optical systems
 the technological demand for a client is readily available.} 
 An additional feature of measurement-only BQC 
is the device-independent \chia{blindness}: even if Alice owns a malicious  device, probably bought from another company, no information is transmitted to Bob because of the no-signaling principle.
\chia{ In our case, we obtain device independence regarding blindness; recently, the concept of device independence has also been generalized to verifiability
~\cite{Hajdusek, Gheorghiu}. }
 Furthermore the full-fledged quantum computer \chia{can be based on } 
any model of MBQC 
(see Ref.\cite{MABQC} for more detailed discussion). 
 
The concept of secure quantum computing opened-up feasible verification methods ~\cite{FK, BarzNP,topoveri}  where Alice can test whether Bob is performing the computation correctly. 
It was shown that the verification is possible for the original protocol~\cite{FK, BarzNP} as well as for the measurement-only protocol ~\cite{topoveri}. 
The central idea in these 
protocols is that Alice secretly
hides some "trap" qubits in the resource state.   This fundamentally reduces to the situation where Alice tries 
to verify,  Bob's quantum resource with a minimal set of measurements \cite{Toth2005, Greganti14, Weinfurter2015}.
If Bob deviates from the correct protocol, he changes the states of the traps,
and if Alice detects the change of any trap, she can detect Bob's malicious behaviour and abort the computation. 
\rev{The security corresponds to the probability that Alice does not accept the results received by a cheating Bob.  }

 In measurement-only BQC
 the trap qubits are  randomly prepared and placed via measurements 
 by Alice within the computation resource and are associated to qubit states in $Z$ and $X$ basis (corresponding to Pauli operators $\sigma_z$ and $\sigma_x$ respectively). When Alice receives the traps, she measures the qubits in the respective basis.

\chia{For the special case of a four-qubit linear cluster state a verification protocol with only two different trap measurements exists.
This protocol runs as follows: 
\begin{itemize}
\item[1.]
Bob prepares the four-qubit linear \rev{cluster} state, and sends each qubit one by one to Alice.
\item[2.]
Alice chooses one of the two tests randomly below:
\begin{itemize}
\item[(a)]
Alice measures qubits 1 and 3 in the $Z$ basis, and qubits 2 and 4 in the $X$ basis.
\item[(b)]
Alice measures qubits 2 and 4 in the $Z$ basis, and qubits 1 and 3 in the $X$ basis.
\end{itemize}
\end{itemize}
If she chooses the option (a), qubits 2 and 4 becomes trap qubits.
If any trap qubit is changed, i.e. she gets not the expected result, then 
she detects Bob's malicious behavior. We call this option (1,3) test.
On the other hand, if she chooses the option (b), qubits 1 and 3 become trap qubits,
and she can check those. We call this option (2,4) test.
We show now that  Bob has to prepare the exact four-qubit linear cluster  state in order to pass all Alice's trap tests in the limit of $n$ repetitions where $n$ tends toward infinity. }
\chia{In the original verification protocols \cite{topoveri, FK}, it is shown that the probability that Alice is fooled by Bob can be exponentially small, by using quantum error correcting codes.}

\rev{Here, let us show the case without quantum error correcting code, which leads to a  probability of accepting a wrong outcome to be polynomially small. We want to point out that this probability can be minimized to become exponentially small by exploiting  standard error amplification techniques ~\cite{Kitaev}  via repeating the computation a number of times proportional to $n$.}
\chia{  Bob can generate any state, but in order to pass the (1,3) test,
Bob has to prepare the state
\begin{equation}
\begin{split}
|\Psi\rangle\equiv
\frac{1}{2}\Big(
|0+0+\rangle|a_1\rangle
+e^{i\theta_2}|0-1-\rangle|a_2\rangle + \\
+e^{i\theta_3}|1-0+\rangle|a_3\rangle  
+e^{i\theta 4}|1+1-\rangle|a_4\rangle
\Big)
\end{split}
\end{equation}
where $\{|a_j\rangle\}$ are certain states of Bob's ancilla,
which are normalized $|\langle a_j|a_j\rangle|^2=1$, but not necesarily mutually orthogonal.  We consider ancilla states, since Bob could prepare a larger system and keep a subsystem. 
Since Bob does not know which option Alice takes,
this state also has to pass the other Alice's test, i.e., (2,4) test.
Assume that Alice gets the result $(0,0)$ when she measures
qubits 2 and 4.
Then the state after the measurement becomes
\begin{equation}
\begin{split}
|\Psi'\rangle\equiv
\frac{1}{2}\Big(
|0\rangle_1|0\rangle_3|a_1\rangle
+e^{i\theta_2}|0\rangle_1|1\rangle_3|a_2\rangle + \\
+e^{i\theta_3}|1\rangle_1|0\rangle_3|a_3\rangle 
+e^{i\theta 4}|1\rangle_1|1\rangle_3|a_4\rangle
\Big).
\end{split}
\end{equation}
In order to pass the (2,4) test (again in the limit of $n$ repetitions), this state must be
$|+\rangle_1|+\rangle_3|b\rangle$ for a certain state $|b\rangle$.
This means that, first, the reduced density operator
of $|\Psi'\rangle$ for Bob's ancilla
\begin{eqnarray*}
\rho=\frac{1}{4}\sum_{j=1}^4|a_j\rangle\langle a_j|
\end{eqnarray*}
must have rank 1, which leads to $|a_j\rangle=|a_k\rangle$ (up to a phase factor).
Now the state is
\begin{eqnarray*}
\frac{1}{2}\Big(
|0\rangle_1|0\rangle_3
+e^{i\theta_2'}|0\rangle_1|1\rangle_3
+e^{i\theta_3'}|1\rangle_1|0\rangle_3
+e^{i\theta 4'}|1\rangle_1|1\rangle_3\Big)|a_1\rangle.
\end{eqnarray*}
Next, in order for this state to be $|+\rangle_1|+\rangle_3|b\rangle$,
$\theta_j'=0$ for all $j=2,3,4$.\\
Therefore, repeating both tests Alice verifies that Bob has the exact four-qubit cluster state, except for a small probability of undetected cheating.

In a general case, Alice can  choose to use the resource state for either verification or computation.  Increasing the number of verifications per computation provides a higher level of security at the cost of efficiency.
The probability of undetected errors, i. e.  Bob cheats in the computation and not in the verification,  is linearly bounded \rev{as in } ~\cite{BarzNP}.
\rev{In Ref.~\cite{BarzNP} Alice is assumed to send qubits, whereas here Alice measures the received qubits. Remarkably the same technique of verification can be applied in both schemes: Alice generates trap qubits via  either choosing Bob's measurement settings or by directly performing the measurements at her side. This allows for the same analysis,  discussed already in Ref. ~\cite{BarzNP}.}
In our case,  Alice can randomly choose between the two verification options and a regular computation on a four-qubit linear cluster.
\mc{Ref.~\cite{Hayashi} describes the asymptotic behavior of the scaling for linear cluster states of increasing length. }
} 

\chia{It is worth to note that in Ref.~\cite{BFK}, a random-number generator is necessary for the blindness, whereas in measurement-only BQC \cite{MABQC}, no random-number generator is required for Alice to guarantee the blindness. If we add the option of the verification, both protocols require random-number generators, since Alice has to randomly place trap qubits. Nevertheless the use of quantum random numbers is nowadays accessible at the consumer grade ~\cite{Sanguinetti}.} \\


\textit{Experiment and results.} -- 
We practically realize a proof-of-principle implementation of the protocol using photons, computing two-qubit entangling gates and verifying two single trap qubits. In contrast to the proposed theoretical scheme \cite{topoveri}, where traps are hidden within the computation resource, our experiment exploits a four-qubit cluster state either for a computation or a verification run, due to the number of available qubits.

\begin{figure}[b]
\begin{center}
\includegraphics[width=0.47\textwidth]{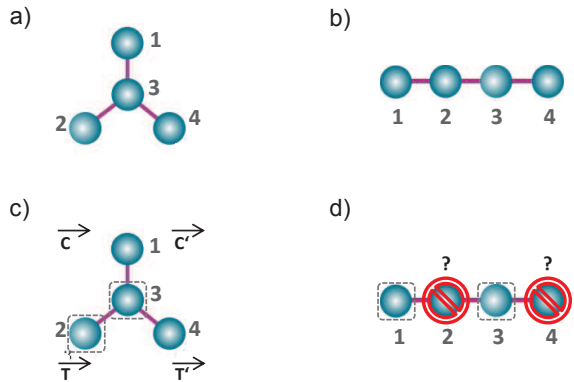}
\end{center} \captionsetup{justification=raggedright}
\caption{
Bob's quantum resources: four-qubit (a) star  and (b) linear  cluster states.  (c) Implementation of an entangling gate on a four-qubit star cluster.  $C$ ($C'$) and $T$ ($T'$) are the control and target input (output) logical qubits.
(d) Verification (1,3)-test on a linear cluster state.  
 The dashed frame corresponds to a single-qubit measurement, whereas the red frame represents a trap.  Verification (2,4)-test is equivalent under exchange of trap and measured qubits.
} 
\label{Cluster}
\end{figure}

The four-qubit resource for measurement-only BQC
, (Fig.~\ref{Cluster} a, b),  is produced in Bob's laboratory via a photonic set-up in a so-called "railway-crossing configuration" (see Fig.~\ref{setup}). A double spontaneous parametric down conversion process allows to generate two pairs of polarization entangled  photons. Interferometers with polarizing beam splitters (PBSs) entangle the four photons. Additional half-wave plates (HWPs) on both pairs directions enable the generation of  different graph states. 
The scheme has been already exploited in several other works  \mc{to create four-qubit linear cluster states and  states that can be obtained from them via local complementations
(see e.g. ~\cite{Barz,Barz2014}). }  
Here we  focus on the generation of a four-qubit star cluster state 
and a four-qubit linear cluster state, respectively (Fig.\ref{Cluster}.a and b): 
\begin{gather}
|C_{star} \rangle= \frac{1}{\sqrt{2}}(|++0+\rangle +|--1-\rangle)_{1234}, \\
|C_{lin} \rangle= \frac{1}{2}(|0+0+\rangle + |0-1-\rangle + |1-0+\rangle + |1+1-\rangle)_{1234},
\end{gather}
where $|\pm\rangle=(|0\rangle \pm |1\rangle)/\sqrt{2}$ are the eigenstates of the Pauli operator $\sigma_x=X$. 
\chia{Remarkably these cluster states belong to different classes of entanglement. 

The four-qubit star cluster was generated within this setup only recently (see \cite{Greganti14} for details) and now exploited for quantum information computing. 
Switching between the two entangled classes involves: preparing specific Bell states at each SPDC process, different photonic interferences between the two pairs of photons,  precise wavelength-scale alignment, and, therefore, high stability. }

Alice's laboratory consists of four HWPs, 
four quarter-wave plates (QWPs), four PBSs and eight single photon counting detectors (APDs) in order to speed up the data acquisition (four APDs are sufficient to measure all possible polarization-basis of four-qubit state) and proceed to a complete  analysis using quantum state tomography (QST)~\cite{Munro}.  
The connecting quantum channel from Bob to Alice is achieved by four single mode fibers,  which carry the  photonic qubits. We reconstructed through over-complete QST the density matrix of the two four-qubit resources, obtaining  fidelities of the state with respect to the ideal star cluster and linear cluster of $F=0.731\pm 0.008$ and $F=0.676\pm 0.007$ (under local unitary operations), respectively (see SI for the density matrix histograms).\\ 

\textit{Computation} -- 
The four-qubit star and linear cluster states are the minimal resources for one-way computation, since the full universal  set of gates 
can be reproduced \cite{Raussendorf2001}. 
 This has been already demonstrated in few works \cite{Walther05, Prevedel07, Chen07, Vallone08, Tokunaga08, BellB2013}. In this work we reproduce different two-qubit 
 entangling gates using the star cluster, in 
order to validate the computation from Alice's device. 
An entangling gate is performed using a star cluster,  
where qubits 1 and 2 are acting as input control ($C$) and target ($T$) qubits, respectively, whereas qubits 1 and 4 present the output control ($C'$) and target ($T'$) qubits,  respectively, as shown in  Fig.\ref{Cluster}c.  
 Different combinations of measurement bases for qubit 2 and 3 enable to create entanglement between the output qubits. Detailed analysis for some entangling gates are reported in the SI. 
 In table \ref{fidCNOT} we present, as example,  the  results related to measuring qubit 2 and 3 in the $Y_2X_3$ basis (where $Y$ is the  Pauli $\sigma_y$), which corresponds to implementing a Controlled NOT (CNOT) gate for input state $|++_i\rangle$ and $|+-_i\rangle$ (where $|+_i\rangle= (|0\rangle + i |1\rangle)/\sqrt{2}$ and $|-_i\rangle= (|0\rangle - i |1\rangle)/\sqrt{2}$ are the eigenstates of $Y$) up to local unitary operations. The two-qubit output states are  analysed through two-qubit QST with acquisition-time of $600s$ per measurement setting. The corresponding uncertanties are due to Poissonian counting statistics and represent only a lower bound for the errors.\\
\begin{table}[h]
  \centering
   \begin{tabular}{ >{\centering}m{1.2cm} >{\centering} m{3.7cm} m{2.4cm}<{\centering}}
   \hline
    $\mathbf{s_2} \mathbf{s_3} $ &  \textbf{Ideal Output State} & \textbf{Fidelity} \\  
   \hline
    $00$    & $(|0+_i\rangle +i|1-_i\rangle)_{14}/\sqrt{2}$ & F=0.87 $\pm$ 0.03 \\
      $01$    & $(|0+_i\rangle +i|1-_i\rangle)_{14}/\sqrt{2}$ & F=0.74 $\pm$ 0.04 \\
      $10$    & $(|0-_i\rangle -i|1+_i\rangle)_{14}/\sqrt{2}$ & F=0.77 $\pm$ 0.03 \\
        $11$    & $(|0-_i\rangle -i|1+_i\rangle)_{14}/\sqrt{2}$ & F=0.77 $\pm$ 0.04 \\
    \hline
   \end{tabular} \captionsetup{justification=raggedright}
\caption{Results from measuring qubit 2 and qubit 3 of the star cluster onto $Y_2X_3$, which corresponds to a CNOT gate on states $|++_{i}\rangle$ and $|+-_i\rangle$ up to a $(Z_1Z_4)^{s_3+1}$, where $s_2$ and $s_3$ are the measurement outcomes.  
The fidelities of the tomographic reconstructed two-qubit state with respect of the ideal state are reported. }
\label{fidCNOT}
 \end{table}

\begin{figure}[h]
\begin{center}
\includegraphics[width=0.49\textwidth]{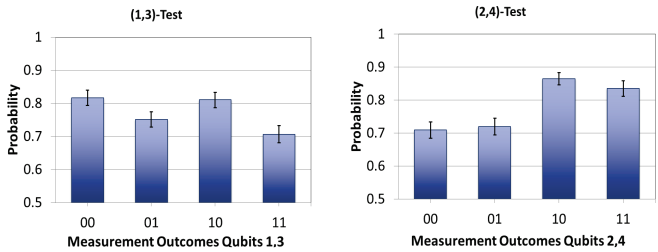}
\end{center} \captionsetup{justification=raggedright}
\caption{Probability that Alice receives the expected outcomes for the (1,3)-test  and (2,4)-test  on a linear cluster state. According to the measurement outcomes of the non-trap qubits (shown on the 	abscissa) we report the probability that measurements on each of Alice's trap qubits return the expected result (i.e. Alice trusts the state Bob sent).
} 
\label{trap}
\end{figure}

  \textit{Verification} -- 
\chia{The four-qubit linear cluster allows verifying computation with only two different trap measurements and is especially suited for the verification protocol as described in the theory section. 
 We present the results for this state along with the results for a 
four-qubit star cluster state. As has been shown in ~\cite{BarzNP}, the 
probability that Alice is fooled by Bob is bounded in such a setting.\\


For the case of the four-qubit linear cluster state we implement the (1,3) and (2,4) tests by having Alice choosing the respective basis. }Per trap we analyze the measurement outcomes  in order to 
 quantify the probabilities that Alice obtains the correct state, 
 see Fig.\ref{trap}. 
 For single trap the results are within the values $[ 0.74 \pm 0.03, 0.98 \pm 0.01 ]$, where the range is due to unbalanced phase noises in the set-up. 
 Each of Alice's measures  has an acquisition-time of $1h$  to decrease the uncertainty. 
Alice verified the resource with non-ideal probability,  
due to experimental imperfections of the set-up, which are present during the generation of the four-qubit resource as already seen from the full QST fidelity of the state. 

\begin{figure}[b]
\begin{center}
\includegraphics[width=0.49\textwidth]{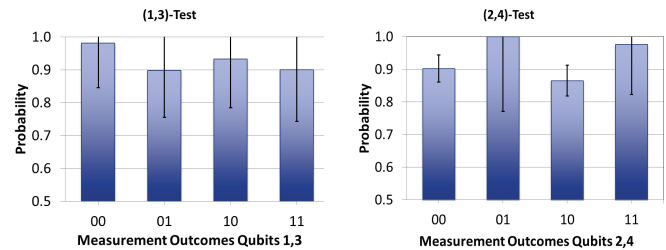}
\end{center} \captionsetup{justification=raggedright}
\caption{ \chia{Probability that Alice receives the expected outcomes for the two verification tests on a star cluster. According to the measurement outcomes of the non-trap qubits (shown on the 	abscissa) we report the probability that measurements on each of Alice's trap qubits return an expected result (i.e. Alice trusts the state Bob sent).}} 
\label{test1Star}
\end{figure}

\chia{Additionally, we performed the verification protocol on a four-qubit star cluster state, which we used before to implement entangling gates. 
We report the two trap tests performed on the star cluster state, equivalently to the linear cluster case, see Fig.\ref{test1Star}.  
 In this case in order to get two trap qubits each time Alice measures  $Z_1Z_3$ (expecting trap qubits in $X$ basis) for (1,3)-test,   and $Z_2X_4$ (expecting one trap qubit in $X$ basis and one in $Z$) for (2,4)-test.} \chia{The single probabilities of individual trap qubits, corresponding to Alice’s expected results, and according to
respective measurement outcomes of the non-trap qubits are in the range $[ 0.90 \pm 0.04, 1.00 -0.16 ]$ with an acquisition time of $600s$ per single measurement (see SI for details).   
  The imbalance of the obtained probabilities with respect to the quantum state fidelity is due to asymmetric noise.  }


\chia{Here we want to point out that the small increased value of fidelity of the star cluster, with respect to the linear cluster, leads to significant improvements for the verification results.}\\ 

\textit{Discussion.} -- 
The use of four-qubit photonic cluster states allows us  to prove the feasibility of two-party measurement-only BQC 
in current quantum optics laboratories. Nevertheless the demonstration can be expanded to several quantum systems and other MBQC models. 
In the photonic case, we want to emphasize that just one HWP, one QWP, one PBS, and one APD would be sufficient for Alice to measure every qubit received from Bob and consequently to implement  computation and verification. The only additional requirement in Bob's laboratory would be a time-delay multiplexer (such as the one used in \cite{Collins}) or a delay line in each photon's path (such as in \cite{Megidish12}).   
The tomographic reconstruction of Bob's resource state,  as we did in our experiment, is in fact superfluous for Alice's computation, since already from the single-qubit measurement she can verify Bob's state.  The quantum power required for Alice is then restricted to measuring the state of the qubits.  
It is important to note the high losses, either due to low detection efficiencies or imperfect quantum channels, would break Alice's computation.  
However, the threshold for losses can be increased by using fault tolerant MBQC models, which are robust against errors and losses \cite{Barrett10}, 
  and besides,   
detection devices with almost unit efficiencies are now available \cite{Lita08, Marsili13}.\\
In conclusion the demonstrated protocol constitutes a  step further to more  realistic secure quantum computing models. \\

\acknowledgements
This work was supported by the European Commission, Initial Training Network PICQUE (No. 608062),
QUILMI (No.295293), EQUAM (No. 323714),  GRASP (No.
613024), and the Vienna Center for Quantum Science and Technology (VCQ), the Austrian Science Fund (FWF) through START (No. Y585-N20) and the doctoral programme CoQuS, the Vienna Science and Technology Fund
(WWTF) under grant ICT12-041, and the Air Force Office of Scientific Research, Air Force Material Command, United States Air Force, under grant number FA8655-11-1-3004, and the Tenure Track System by MEXT Japan and KAKENHI 26730003 and 15H00850 by JSPS.


\bibliographystyle{apsrev}
\bibliography{_mybib5}

\begin{thebibliography}{39}
\expandafter\ifx\csname natexlab\endcsname\relax\def\natexlab#1{#1}\fi
\expandafter\ifx\csname bibnamefont\endcsname\relax
  \def\bibnamefont#1{#1}\fi
\expandafter\ifx\csname bibfnamefont\endcsname\relax
  \def\bibfnamefont#1{#1}\fi
\expandafter\ifx\csname citenamefont\endcsname\relax
  \def\citenamefont#1{#1}\fi
\expandafter\ifx\csname url\endcsname\relax
  \def\url#1{\texttt{#1}}\fi
\expandafter\ifx\csname urlprefix\endcsname\relax\def\urlprefix{URL }\fi
\providecommand{\bibinfo}[2]{#2}
\providecommand{\eprint}[2][]{\url{#2}}

\bibitem[{\citenamefont{Broadbent et~al.}(2009)\citenamefont{Broadbent,
  Fitzsimons, and Kashefi}}]{BFK}
\bibinfo{author}{\bibfnamefont{A.}~\bibnamefont{Broadbent}},
  \bibinfo{author}{\bibfnamefont{J.}~\bibnamefont{Fitzsimons}},
  \bibnamefont{and} \bibinfo{author}{\bibfnamefont{E.}~\bibnamefont{Kashefi}},
  in \emph{\bibinfo{booktitle}{Proceedings of the 50th Annual Symposium on
  Foundations of Computer Science}} (\bibinfo{year}{2009}), pp.
  \bibinfo{pages}{517--526}.

\bibitem[{\citenamefont{Morimae and Fujii}(2013)}]{MABQC}
\bibinfo{author}{\bibfnamefont{T.}~\bibnamefont{Morimae}} \bibnamefont{and}
  \bibinfo{author}{\bibfnamefont{K.}~\bibnamefont{Fujii}},
  \bibinfo{journal}{Phys. Rev. A} \textbf{\bibinfo{volume}{87}},
  \bibinfo{pages}{050301} (\bibinfo{year}{2013}).

\bibitem[{\citenamefont{Fitzsimons and Kashefi}(2012)}]{FK}
\bibinfo{author}{\bibfnamefont{J.}~\bibnamefont{Fitzsimons}} \bibnamefont{and}
  \bibinfo{author}{\bibfnamefont{E.}~\bibnamefont{Kashefi}},
  \bibinfo{journal}{arXiv:1203.5217}  (\bibinfo{year}{2012}).

\bibitem[{\citenamefont{Dunjko et~al.}(2012)\citenamefont{Dunjko, Kashefi, and
  Leverrier}}]{Vedran}
\bibinfo{author}{\bibfnamefont{V.}~\bibnamefont{Dunjko}},
  \bibinfo{author}{\bibfnamefont{E.}~\bibnamefont{Kashefi}}, \bibnamefont{and}
  \bibinfo{author}{\bibfnamefont{A.}~\bibnamefont{Leverrier}},
  \bibinfo{journal}{Phys. Rev. Lett.} \textbf{\bibinfo{volume}{108}},
  \bibinfo{pages}{200502} (\bibinfo{year}{2012}).

\bibitem[{\citenamefont{Morimae et~al.}(2015)\citenamefont{Morimae, Dunjko, and
  Kashefi}}]{AKLTblind}
\bibinfo{author}{\bibfnamefont{T.}~\bibnamefont{Morimae}},
  \bibinfo{author}{\bibfnamefont{V.}~\bibnamefont{Dunjko}}, \bibnamefont{and}
  \bibinfo{author}{\bibfnamefont{E.}~\bibnamefont{Kashefi}},
  \bibinfo{journal}{Quantum Information and Computation}
  \textbf{\bibinfo{volume}{15}}, \bibinfo{pages}{0200} (\bibinfo{year}{2015}).

\bibitem[{\citenamefont{Morimae and Fujii}(2012)}]{topoblind}
\bibinfo{author}{\bibfnamefont{T.}~\bibnamefont{Morimae}} \bibnamefont{and}
  \bibinfo{author}{\bibfnamefont{K.}~\bibnamefont{Fujii}},
  \bibinfo{journal}{Nat. Commun.} \textbf{\bibinfo{volume}{3}},
  \bibinfo{pages}{1036} (\bibinfo{year}{2012}).

\bibitem[{\citenamefont{Morimae}(2012)}]{CVblind}
\bibinfo{author}{\bibfnamefont{T.}~\bibnamefont{Morimae}},
  \bibinfo{journal}{Phys. Rev. Lett.} \textbf{\bibinfo{volume}{109}},
  \bibinfo{pages}{230502} (\bibinfo{year}{2012}).

\bibitem[{\citenamefont{Giovannetti et~al.}(2013)\citenamefont{Giovannetti,
  Maccone, Morimae, and Rudolph}}]{Lorenzo}
\bibinfo{author}{\bibfnamefont{V.}~\bibnamefont{Giovannetti}},
  \bibinfo{author}{\bibfnamefont{L.}~\bibnamefont{Maccone}},
  \bibinfo{author}{\bibfnamefont{T.}~\bibnamefont{Morimae}}, \bibnamefont{and}
  \bibinfo{author}{\bibfnamefont{T.}~\bibnamefont{Rudolph}},
  \bibinfo{journal}{Phys. Rev. Lett.} \textbf{\bibinfo{volume}{111}},
  \bibinfo{pages}{230501} (\bibinfo{year}{2013}).

\bibitem[{\citenamefont{Mantri et~al.}(2013)\citenamefont{Mantri,
  P\'erez-Delgado, and Fitzsimons}}]{Joe_intern}
\bibinfo{author}{\bibfnamefont{A.}~\bibnamefont{Mantri}},
  \bibinfo{author}{\bibfnamefont{C.}~\bibnamefont{P\'erez-Delgado}},
  \bibnamefont{and}
  \bibinfo{author}{\bibfnamefont{J.}~\bibnamefont{Fitzsimons}},
  \bibinfo{journal}{Phys. Rev. Lett.} \textbf{\bibinfo{volume}{111}},
  \bibinfo{pages}{230502} (\bibinfo{year}{2013}).

\bibitem[{\citenamefont{Sueki et~al.}(2013)\citenamefont{Sueki, Koshiba, and
  Morimae}}]{Sueki}
\bibinfo{author}{\bibfnamefont{T.}~\bibnamefont{Sueki}},
  \bibinfo{author}{\bibfnamefont{T.}~\bibnamefont{Koshiba}}, \bibnamefont{and}
  \bibinfo{author}{\bibfnamefont{T.}~\bibnamefont{Morimae}},
  \bibinfo{journal}{Phys. Rev. A} \textbf{\bibinfo{volume}{87}},
  \bibinfo{pages}{060301} (\bibinfo{year}{2013}).

\bibitem[{\citenamefont{Li et~al.}(2014)\citenamefont{Li, Chan, Wu, and
  Wen}}]{tri}
\bibinfo{author}{\bibfnamefont{Q.}~\bibnamefont{Li}},
  \bibinfo{author}{\bibfnamefont{W.~H.} \bibnamefont{Chan}},
  \bibinfo{author}{\bibfnamefont{C.}~\bibnamefont{Wu}}, \bibnamefont{and}
  \bibinfo{author}{\bibfnamefont{Z.}~\bibnamefont{Wen}},
  \bibinfo{journal}{Phys. Rev. A} \textbf{\bibinfo{volume}{89}},
  \bibinfo{pages}{040302} (\bibinfo{year}{2014}).

\bibitem[{\citenamefont{Barz et~al.}(2012)\citenamefont{Barz, Kashefi,
  Broadbent, Fitzsimons, Zeilinger, and Walther}}]{Barz}
\bibinfo{author}{\bibfnamefont{S.}~\bibnamefont{Barz}},
  \bibinfo{author}{\bibfnamefont{E.}~\bibnamefont{Kashefi}},
  \bibinfo{author}{\bibfnamefont{A.}~\bibnamefont{Broadbent}},
  \bibinfo{author}{\bibfnamefont{J.}~\bibnamefont{Fitzsimons}},
  \bibinfo{author}{\bibfnamefont{A.}~\bibnamefont{Zeilinger}},
  \bibnamefont{and} \bibinfo{author}{\bibfnamefont{P.}~\bibnamefont{Walther}},
  \bibinfo{journal}{Science} \textbf{\bibinfo{volume}{335}},
  \bibinfo{pages}{303} (\bibinfo{year}{2012}).

\bibitem[{\citenamefont{Barz et~al.}(2013)\citenamefont{Barz, Fitzsimons,
  Kashefi, and Walther}}]{BarzNP}
\bibinfo{author}{\bibfnamefont{S.}~\bibnamefont{Barz}},
  \bibinfo{author}{\bibfnamefont{J.~F.} \bibnamefont{Fitzsimons}},
  \bibinfo{author}{\bibfnamefont{E.}~\bibnamefont{Kashefi}}, \bibnamefont{and}
  \bibinfo{author}{\bibfnamefont{P.}~\bibnamefont{Walther}},
  \bibinfo{journal}{Nature Physics} \textbf{\bibinfo{volume}{9}},
  \bibinfo{pages}{727} (\bibinfo{year}{2013}).

\bibitem[{\citenamefont{Fisher et~al.}(2014)\citenamefont{Fisher, Broadbent,
  Shalm, Yan, Lavoie, Prevedel, Jennewein, and Resch}}]{Fisher14}
\bibinfo{author}{\bibfnamefont{K.}~\bibnamefont{Fisher}},
  \bibinfo{author}{\bibfnamefont{A.}~\bibnamefont{Broadbent}},
  \bibinfo{author}{\bibfnamefont{L.}~\bibnamefont{Shalm}},
  \bibinfo{author}{\bibfnamefont{Z.}~\bibnamefont{Yan}},
  \bibinfo{author}{\bibfnamefont{J.}~\bibnamefont{Lavoie}},
  \bibinfo{author}{\bibfnamefont{R.}~\bibnamefont{Prevedel}},
  \bibinfo{author}{\bibfnamefont{T.}~\bibnamefont{Jennewein}},
  \bibnamefont{and} \bibinfo{author}{\bibfnamefont{K.}~\bibnamefont{Resch}},
  \bibinfo{journal}{Nature Communications} \textbf{\bibinfo{volume}{5}},
  \bibinfo{pages}{3074} (\bibinfo{year}{2014}).

\bibitem[{\citenamefont{Morimae}(2014)}]{topoveri}
\bibinfo{author}{\bibfnamefont{T.}~\bibnamefont{Morimae}},
  \bibinfo{journal}{Phys. Rev. A} \textbf{\bibinfo{volume}{89}},
  \bibinfo{pages}{060302} (\bibinfo{year}{2014}).

\bibitem[{\citenamefont{Popescu and Rohrlich}(1994)}]{PR}
\bibinfo{author}{\bibfnamefont{S.}~\bibnamefont{Popescu}} \bibnamefont{and}
  \bibinfo{author}{\bibfnamefont{D.}~\bibnamefont{Rohrlich}},
  \bibinfo{journal}{Foundations of Physics} \textbf{\bibinfo{volume}{24}},
  \bibinfo{pages}{379} (\bibinfo{year}{1994}), ISSN \bibinfo{issn}{0015-9018}.

\bibitem[{\citenamefont{Raussendorf and Briegel}(2001)}]{Raussendorf2001}
\bibinfo{author}{\bibfnamefont{R.}~\bibnamefont{Raussendorf}} \bibnamefont{and}
  \bibinfo{author}{\bibfnamefont{H.}~\bibnamefont{Briegel}},
  \bibinfo{journal}{Phys. Rev. Lett.} \textbf{\bibinfo{volume}{86}},
  \bibinfo{pages}{5188} (\bibinfo{year}{2001}).

\bibitem[{\citenamefont{Briegel et~al.}(2009)\citenamefont{Briegel, Browne,
  D{\"u}r, Raussendorf, and Van~den Nest}}]{Briegel2009}
\bibinfo{author}{\bibfnamefont{H.}~\bibnamefont{Briegel}},
  \bibinfo{author}{\bibfnamefont{D.~E.} \bibnamefont{Browne}},
  \bibinfo{author}{\bibfnamefont{W.}~\bibnamefont{D{\"u}r}},
  \bibinfo{author}{\bibfnamefont{R.}~\bibnamefont{Raussendorf}},
  \bibnamefont{and} \bibinfo{author}{\bibfnamefont{M.}~\bibnamefont{Van~den
  Nest}}, \bibinfo{journal}{Nature Phys.} \textbf{\bibinfo{volume}{5}},
  \bibinfo{pages}{19} (\bibinfo{year}{2009}), ISSN \bibinfo{issn}{1745-2473}.

\bibitem[{\citenamefont{Hajdusek et~al.}(2015)\citenamefont{Hajdusek,
  P\'erez-Delgado, and Fitzsimons}}]{Hajdusek}
\bibinfo{author}{\bibfnamefont{M.}~\bibnamefont{Hajdusek}},
  \bibinfo{author}{\bibfnamefont{C.}~\bibnamefont{P\'erez-Delgado}},
  \bibnamefont{and}
  \bibinfo{author}{\bibfnamefont{J.}~\bibnamefont{Fitzsimons}},
  \bibinfo{journal}{arXiv:1502.02563v1}  (\bibinfo{year}{2015}).

\bibitem[{\citenamefont{Gheorghiu et~al.}(2015)\citenamefont{Gheorghiu,
  Kashefi, and Wallden}}]{Gheorghiu}
\bibinfo{author}{\bibfnamefont{A.}~\bibnamefont{Gheorghiu}},
  \bibinfo{author}{\bibfnamefont{E.}~\bibnamefont{Kashefi}}, \bibnamefont{and}
  \bibinfo{author}{\bibfnamefont{P.}~\bibnamefont{Wallden}},
  \bibinfo{journal}{arXiv:1502.02571v2}  (\bibinfo{year}{2015}).

\bibitem[{\citenamefont{T\'oth and G\"uhne}(2005)}]{Toth2005}
\bibinfo{author}{\bibfnamefont{G.}~\bibnamefont{T\'oth}} \bibnamefont{and}
  \bibinfo{author}{\bibfnamefont{O.}~\bibnamefont{G\"uhne}},
  \bibinfo{journal}{Phys. Rev. A} \textbf{\bibinfo{volume}{72}},
  \bibinfo{pages}{022340} (\bibinfo{year}{2005}).

\bibitem[{\citenamefont{Greganti et~al.}(2015)\citenamefont{Greganti, Roehsner,
  Barz, Waegell, and Walther}}]{Greganti14}
\bibinfo{author}{\bibfnamefont{C.}~\bibnamefont{Greganti}},
  \bibinfo{author}{\bibfnamefont{M.~C.} \bibnamefont{Roehsner}},
  \bibinfo{author}{\bibfnamefont{S.}~\bibnamefont{Barz}},
  \bibinfo{author}{\bibfnamefont{M.}~\bibnamefont{Waegell}}, \bibnamefont{and}
  \bibinfo{author}{\bibfnamefont{P.}~\bibnamefont{Walther}},
  \bibinfo{journal}{Phys. Rev. A, 91 022325}  (\bibinfo{year}{2015}).

\bibitem[{\citenamefont{Knips et~al.}(2014)\citenamefont{Knips, Schwemmer,
  Klein, {Wie{\'s}niak}, and Weinfurter}}]{Weinfurter2015}
\bibinfo{author}{\bibfnamefont{L.}~\bibnamefont{Knips}},
  \bibinfo{author}{\bibfnamefont{C.}~\bibnamefont{Schwemmer}},
  \bibinfo{author}{\bibfnamefont{N.}~\bibnamefont{Klein}},
  \bibinfo{author}{\bibfnamefont{M.}~\bibnamefont{{Wie{\'s}niak}}},
  \bibnamefont{and}
  \bibinfo{author}{\bibfnamefont{H.}~\bibnamefont{Weinfurter}},
  \bibinfo{journal}{arXiv:1412.5881}  (\bibinfo{year}{2014}).

\bibitem[{\citenamefont{Kitaev et~al.}(2002)\citenamefont{Kitaev, Shen, and
  Vyalyi}}]{Kitaev}
\bibinfo{author}{\bibfnamefont{A.~Y.} \bibnamefont{Kitaev}},
  \bibinfo{author}{\bibfnamefont{A.~H.} \bibnamefont{Shen}}, \bibnamefont{and}
  \bibinfo{author}{\bibfnamefont{M.~N.} \bibnamefont{Vyalyi}},
  \emph{\bibinfo{title}{Classical and Quantum Computation}}
  (\bibinfo{publisher}{American Mathematical Society},
  \bibinfo{address}{Boston, MA, USA}, \bibinfo{year}{2002}), ISBN
  \bibinfo{isbn}{0821832298}.

\bibitem[{\citenamefont{Hayashi and Morimae}(2015)}]{Hayashi}
\bibinfo{author}{\bibfnamefont{M.}~\bibnamefont{Hayashi}} \bibnamefont{and}
  \bibinfo{author}{\bibfnamefont{T.}~\bibnamefont{Morimae}},
  \bibinfo{journal}{arXiv:1505.07535}  (\bibinfo{year}{2015}).

\bibitem[{\citenamefont{Sanguinetti et~al.}(2014)\citenamefont{Sanguinetti,
  Martin, Zbinden, and Gisin}}]{Sanguinetti}
\bibinfo{author}{\bibfnamefont{B.}~\bibnamefont{Sanguinetti}},
  \bibinfo{author}{\bibfnamefont{A.}~\bibnamefont{Martin}},
  \bibinfo{author}{\bibfnamefont{H.}~\bibnamefont{Zbinden}}, \bibnamefont{and}
  \bibinfo{author}{\bibfnamefont{N.}~\bibnamefont{Gisin}},
  \bibinfo{journal}{Phys. Rev. X} \textbf{\bibinfo{volume}{4}},
  \bibinfo{pages}{031056} (\bibinfo{year}{2014}).

\bibitem[{\citenamefont{Barz et~al.}(2014)\citenamefont{Barz, Vasconcelos,
  Greganti, Zwerger, Duer, Briegel, and Walther}}]{Barz2014}
\bibinfo{author}{\bibfnamefont{S.}~\bibnamefont{Barz}},
  \bibinfo{author}{\bibfnamefont{R.}~\bibnamefont{Vasconcelos}},
  \bibinfo{author}{\bibfnamefont{C.}~\bibnamefont{Greganti}},
  \bibinfo{author}{\bibfnamefont{M.}~\bibnamefont{Zwerger}},
  \bibinfo{author}{\bibfnamefont{W.}~\bibnamefont{Duer}},
  \bibinfo{author}{\bibfnamefont{H.}~\bibnamefont{Briegel}}, \bibnamefont{and}
  \bibinfo{author}{\bibfnamefont{P.}~\bibnamefont{Walther}},
  \bibinfo{journal}{arXiv. 1308.5209}  (\bibinfo{year}{2014}).

\bibitem[{\citenamefont{James et~al.}(2001)\citenamefont{James, Kwiat, Munro,
  and White}}]{Munro}
\bibinfo{author}{\bibfnamefont{D.}~\bibnamefont{James}},
  \bibinfo{author}{\bibfnamefont{P.}~\bibnamefont{Kwiat}},
  \bibinfo{author}{\bibfnamefont{W.}~\bibnamefont{Munro}}, \bibnamefont{and}
  \bibinfo{author}{\bibfnamefont{A.}~\bibnamefont{White}},
  \bibinfo{journal}{Phys. Rev. A} \textbf{\bibinfo{volume}{64}},
  \bibinfo{pages}{52312} (\bibinfo{year}{2001}).

\bibitem[{\citenamefont{Walther et~al.}(2005)\citenamefont{Walther, Resch,
  Rudolph, Schenck, Weinfurter, Vedral, Aspelmeyer, and Zeilinger}}]{Walther05}
\bibinfo{author}{\bibfnamefont{P.}~\bibnamefont{Walther}},
  \bibinfo{author}{\bibfnamefont{K.}~\bibnamefont{Resch}},
  \bibinfo{author}{\bibfnamefont{T.}~\bibnamefont{Rudolph}},
  \bibinfo{author}{\bibfnamefont{E.}~\bibnamefont{Schenck}},
  \bibinfo{author}{\bibfnamefont{H.}~\bibnamefont{Weinfurter}},
  \bibinfo{author}{\bibfnamefont{V.}~\bibnamefont{Vedral}},
  \bibinfo{author}{\bibfnamefont{M.}~\bibnamefont{Aspelmeyer}},
  \bibnamefont{and}
  \bibinfo{author}{\bibfnamefont{A.}~\bibnamefont{Zeilinger}},
  \bibinfo{journal}{Nature} \textbf{\bibinfo{volume}{434}},
  \bibinfo{pages}{169} (\bibinfo{year}{2005}).

\bibitem[{\citenamefont{Prevedel et~al.}(2007)\citenamefont{Prevedel, Walther,
  Tiefenbacher, B\"{o}hi, Kaltenbaek, Jennewein, and Zeilinger}}]{Prevedel07}
\bibinfo{author}{\bibfnamefont{R.}~\bibnamefont{Prevedel}},
  \bibinfo{author}{\bibfnamefont{P.}~\bibnamefont{Walther}},
  \bibinfo{author}{\bibfnamefont{F.}~\bibnamefont{Tiefenbacher}},
  \bibinfo{author}{\bibfnamefont{P.}~\bibnamefont{B\"{o}hi}},
  \bibinfo{author}{\bibfnamefont{R.}~\bibnamefont{Kaltenbaek}},
  \bibinfo{author}{\bibfnamefont{T.}~\bibnamefont{Jennewein}},
  \bibnamefont{and}
  \bibinfo{author}{\bibfnamefont{A.}~\bibnamefont{Zeilinger}},
  \bibinfo{journal}{Nature} \textbf{\bibinfo{volume}{445}}, \bibinfo{pages}{65}
  (\bibinfo{year}{2007}).

\bibitem[{\citenamefont{Chen et~al.}(2007)\citenamefont{Chen, Li, Zhang, Chen,
  Goebel, Chen, Mair, and Pan}}]{Chen07}
\bibinfo{author}{\bibfnamefont{K.}~\bibnamefont{Chen}},
  \bibinfo{author}{\bibfnamefont{C.-M.} \bibnamefont{Li}},
  \bibinfo{author}{\bibfnamefont{Q.}~\bibnamefont{Zhang}},
  \bibinfo{author}{\bibfnamefont{Y.-A.} \bibnamefont{Chen}},
  \bibinfo{author}{\bibfnamefont{A.}~\bibnamefont{Goebel}},
  \bibinfo{author}{\bibfnamefont{S.}~\bibnamefont{Chen}},
  \bibinfo{author}{\bibfnamefont{A.}~\bibnamefont{Mair}}, \bibnamefont{and}
  \bibinfo{author}{\bibfnamefont{J.-W.} \bibnamefont{Pan}},
  \bibinfo{journal}{Phys. Rev. Lett.} \textbf{\bibinfo{volume}{99}},
  \bibinfo{eid}{120503} (pages~\bibinfo{numpages}{4}) (\bibinfo{year}{2007}).

\bibitem[{\citenamefont{Vallone et~al.}(2008)\citenamefont{Vallone, Pomarico,
  Martini, and Mataloni}}]{Vallone08}
\bibinfo{author}{\bibfnamefont{G.}~\bibnamefont{Vallone}},
  \bibinfo{author}{\bibfnamefont{E.}~\bibnamefont{Pomarico}},
  \bibinfo{author}{\bibfnamefont{F.~D.} \bibnamefont{Martini}},
  \bibnamefont{and} \bibinfo{author}{\bibfnamefont{P.}~\bibnamefont{Mataloni}},
  \bibinfo{journal}{Phys. Rev. A} \textbf{\bibinfo{volume}{78}},
  \bibinfo{pages}{042335} (\bibinfo{year}{2008}).

\bibitem[{\citenamefont{Tokunaga et~al.}(2008)\citenamefont{Tokunaga,
  Kuwashiro, Yamamoto, Koashi, and Imoto}}]{Tokunaga08}
\bibinfo{author}{\bibfnamefont{Y.}~\bibnamefont{Tokunaga}},
  \bibinfo{author}{\bibfnamefont{S.}~\bibnamefont{Kuwashiro}},
  \bibinfo{author}{\bibfnamefont{T.}~\bibnamefont{Yamamoto}},
  \bibinfo{author}{\bibfnamefont{M.}~\bibnamefont{Koashi}}, \bibnamefont{and}
  \bibinfo{author}{\bibfnamefont{N.}~\bibnamefont{Imoto}},
  \bibinfo{journal}{Phys. Rev. Lett.} \textbf{\bibinfo{volume}{100}},
  \bibinfo{pages}{210501} (\bibinfo{year}{2008}), ISSN
  \bibinfo{issn}{1079-7114}.

\bibitem[{\citenamefont{Bell et~al.}(2013)\citenamefont{Bell, Tame, Clark,
  Nock, Wadsworth, and Rarity}}]{BellB2013}
\bibinfo{author}{\bibfnamefont{B.}~\bibnamefont{Bell}},
  \bibinfo{author}{\bibfnamefont{M.}~\bibnamefont{Tame}},
  \bibinfo{author}{\bibfnamefont{A.}~\bibnamefont{Clark}},
  \bibinfo{author}{\bibfnamefont{R.}~\bibnamefont{Nock}},
  \bibinfo{author}{\bibfnamefont{W.}~\bibnamefont{Wadsworth}},
  \bibnamefont{and} \bibinfo{author}{\bibfnamefont{J.}~\bibnamefont{Rarity}},
  \bibinfo{journal}{New J. Phys.} \textbf{\bibinfo{volume}{15}},
  \bibinfo{pages}{053030} (\bibinfo{year}{2013}).

\bibitem[{\citenamefont{Collins et~al.}(2013)\citenamefont{Collins, Xiong, Rey,
  Vo, He, Shahnia, C.~Reardon, Steel, Clark, and Eggleton}}]{Collins}
\bibinfo{author}{\bibfnamefont{M.}~\bibnamefont{Collins}},
  \bibinfo{author}{\bibfnamefont{C.}~\bibnamefont{Xiong}},
  \bibinfo{author}{\bibfnamefont{I.}~\bibnamefont{Rey}},
  \bibinfo{author}{\bibfnamefont{T.}~\bibnamefont{Vo}},
  \bibinfo{author}{\bibfnamefont{J.}~\bibnamefont{He}},
  \bibinfo{author}{\bibfnamefont{S.}~\bibnamefont{Shahnia}},
  \bibinfo{author}{\bibfnamefont{T.~K.} \bibnamefont{C.~Reardon}},
  \bibinfo{author}{\bibfnamefont{M.}~\bibnamefont{Steel}},
  \bibinfo{author}{\bibfnamefont{A.}~\bibnamefont{Clark}}, \bibnamefont{and}
  \bibinfo{author}{\bibfnamefont{B.}~\bibnamefont{Eggleton}},
  \bibinfo{journal}{Nat. Comm.} \textbf{\bibinfo{volume}{4}},
  \bibinfo{pages}{2582} (\bibinfo{year}{2013}).

\bibitem[{\citenamefont{Megidish et~al.}(2013)\citenamefont{Megidish, Halevy,
  Shacham, Dvir, Dovrat, and Eisenberg}}]{Megidish12}
\bibinfo{author}{\bibfnamefont{E.}~\bibnamefont{Megidish}},
  \bibinfo{author}{\bibfnamefont{A.}~\bibnamefont{Halevy}},
  \bibinfo{author}{\bibfnamefont{T.}~\bibnamefont{Shacham}},
  \bibinfo{author}{\bibfnamefont{T.}~\bibnamefont{Dvir}},
  \bibinfo{author}{\bibfnamefont{L.}~\bibnamefont{Dovrat}}, \bibnamefont{and}
  \bibinfo{author}{\bibfnamefont{H.~S.} \bibnamefont{Eisenberg}},
  \bibinfo{journal}{Phys. Rev. Lett.} \textbf{\bibinfo{volume}{110}},
  \bibinfo{pages}{210403} (\bibinfo{year}{2013}).

\bibitem[{\citenamefont{Barrett and Stace}(2010)}]{Barrett10}
\bibinfo{author}{\bibfnamefont{S.}~\bibnamefont{Barrett}} \bibnamefont{and}
  \bibinfo{author}{\bibfnamefont{T.~M.} \bibnamefont{Stace}},
  \bibinfo{journal}{Phys. Rev. Lett.} \textbf{\bibinfo{volume}{105}},
  \bibinfo{pages}{200502} (\bibinfo{year}{2010}).

\bibitem[{\citenamefont{Lita et~al.}(2008)\citenamefont{Lita, Miller, and
  Nam}}]{Lita08}
\bibinfo{author}{\bibfnamefont{A.~E.} \bibnamefont{Lita}},
  \bibinfo{author}{\bibfnamefont{A.~J.} \bibnamefont{Miller}},
  \bibnamefont{and} \bibinfo{author}{\bibfnamefont{S.~W.} \bibnamefont{Nam}},
  \bibinfo{journal}{Opt. Express} \textbf{\bibinfo{volume}{16}},
  \bibinfo{pages}{3032} (\bibinfo{year}{2008}).

\bibitem[{\citenamefont{Marsili et~al.}(2013)\citenamefont{Marsili, Verma,
  Stern, Harrington, A.~E.~Lita, Vayshenker, Baek, Shaw, Mirin, and
  Nam}}]{Marsili13}
\bibinfo{author}{\bibfnamefont{F.}~\bibnamefont{Marsili}},
  \bibinfo{author}{\bibfnamefont{V.~B.} \bibnamefont{Verma}},
  \bibinfo{author}{\bibfnamefont{J.~A.} \bibnamefont{Stern}},
  \bibinfo{author}{\bibfnamefont{S.}~\bibnamefont{Harrington}},
  \bibinfo{author}{\bibfnamefont{T.~G.} \bibnamefont{A.~E.~Lita}},
  \bibinfo{author}{\bibfnamefont{I.}~\bibnamefont{Vayshenker}},
  \bibinfo{author}{\bibfnamefont{B.}~\bibnamefont{Baek}},
  \bibinfo{author}{\bibfnamefont{M.~D.} \bibnamefont{Shaw}},
  \bibinfo{author}{\bibfnamefont{R.~P.} \bibnamefont{Mirin}}, \bibnamefont{and}
  \bibinfo{author}{\bibfnamefont{S.~W.} \bibnamefont{Nam}},
  \bibinfo{journal}{Nat. Phot.} \textbf{\bibinfo{volume}{7}},
  \bibinfo{pages}{210} (\bibinfo{year}{2013}).

\end{thebibliography}

\end{document}


\title{Supplementary Information \\ ''Demonstration of measurement-only blind quantum computing''}
\author{Chiara Greganti $^1$, Marie-Christine Roehsner$^1$,  Stefanie Barz$^2$, Tomoyuki Morimae$^3$, Philip Walther$^1$}
\affiliation{$^1$~University of Vienna, Faculty of Physics, Austria, $^2$~University of Oxford,  Clarendon Laboratory, UK, $^3$~ASRLD Unit, Gunma University, 1-5-1 Tenjin-cho Kiryu-shi Gunma-ken, 376-0052, Japan}
\maketitle

\tableofcontents

\newpage

\newpage
\section{Analysis of Bob's Resource State}

The tomographic density matrices of the four-qubit star and linear cluster states are shown in Fig.\ref{QST_histo}.
Each measurement setting is acquired for $600s$ and with a four-fold count rate of $0.33Hz$.\\

\begin{figure}[hb]
\begin{center}
\includegraphics[width=0.8\textwidth]{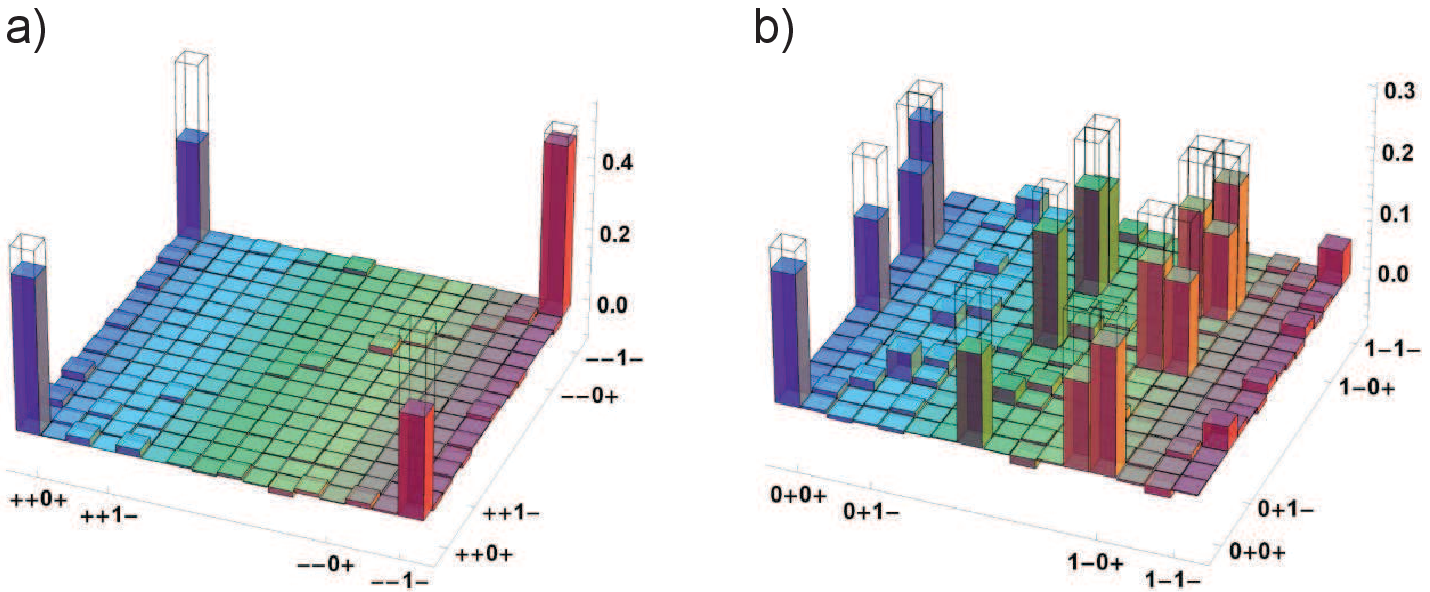}
\end{center} \captionsetup{justification=raggedright}
\caption{ Tomographic reconstructed density  matrices for a four-qubit star ( \textbf{a)} presented on $XXZX$ basis) and linear ( \textbf{b)} presented on $ZXZX$ basis)  cluster states. The fidelities with respect of the ideal states are $F=0.731 \pm 0.008$ and $F=0.676 \pm 0.007$, respectively, under local unitary rotations. } 
\label{QST_histo}
\end{figure}

\section{Analysis of Entangling Gates on Four-qubit Star Cluster State}

We analyse the creation of entanglement in different two-qubit entangling  gates on the four-qubit star cluster state. 
For the general C-gate scheme see in main text Fig.4.c. 
Below we report the fidelities of the output entangled two-qubit states according to different combinations of measurement basis for the qubit 2 and 3 of the star cluster state. \\
$Z$, $X$ and $Y$ are the Pauli operators $\sigma_z$, $\sigma_x$ and $\sigma_y$, respectively.  $s_2$ and $s_3$ are the measurement outcomes for qubit 2 and 3.  $CNOT$ and C-Phase($\pi/2$) gates correspond to two-qubit gate $(|0\rangle_i \langle 0| \otimes I_j + |1\rangle_i \langle 1| \otimes X_j   )$ and $(|0\rangle_i \langle 0| \otimes I_j + |1\rangle_i \langle 1| \otimes R(\pi/2)_j )$, respectively, where $R(\pi/2)$ is a one-qubit rotation of $\pi/2$ around $Z$ axis. $H$ is the Hadamard gate, $H= Z+X$ and H$_y$ is equal to $Z+Y$.\\
The two-qubit output states are  analysed through two-qubit QST with acquisition-time of $600s$ per measurement setting. The corresponding uncertanties are due to Poissonian counting statistics and represent only a lower bound for the errors.

 \begin{table}[h!]
  \centering
   \begin{tabular}{ >{\centering}m{1.2cm} >{\centering} m{3.7cm} m{2.4cm}<{\centering}}
   \hline
    $\mathbf{s_2} \mathbf{s_3} $ &  \textbf{Ideal Output State} & \textbf{Fidelity} \\  
   \hline
    $00$    & $(|00\rangle +|11\rangle)_{14}/\sqrt{2}$ & F=0.76 $\pm$ 0.03 \\
      $01$    & $(|00\rangle +|11\rangle)_{14}/\sqrt{2}$ & F=0.70 $\pm$ 0.03 \\
      $10$    & $(|01\rangle +|10\rangle)_{14}/\sqrt{2}$ & F=0.79 $\pm$ 0.04 \\
        $11$    & $(|01\rangle +|10\rangle)_{14}/\sqrt{2}$ & F=0.68 $\pm$ 0.04 \\
    \hline
   \end{tabular}\captionsetup{justification=raggedright}
\caption{Results from measuring qubit 2 and qubit 3 of the star cluster onto $Z_2X_3$, which corresponds to a CNOT gate on states $|+0\rangle$ and $|+1\rangle$ up to a $(X_4)^{s_3}$.  
The fidelities of the tomographic reconstructed state with respect of the ideal state are reported. }
 \end{table}

 \begin{table}[h!]
  \centering
   \begin{tabular}{ >{\centering}m{1.2cm} >{\centering} m{3.7cm} m{2.4cm}<{\centering}}
   \hline
    $\mathbf{s_2} \mathbf{s_3} $ &  \textbf{Ideal Output State} & \textbf{Fidelity} \\  
   \hline
  $00$    & $(|00\rangle +|11\rangle)_{14}/\sqrt{2}$ & F=0.73 $\pm$ 0.03 \\
      $01$    & $(|00\rangle +|11\rangle)_{14}/\sqrt{2}$ & F=0.73 $\pm$ 0.03 \\
      $10$    & $(|01\rangle +|10\rangle)_{14}/\sqrt{2}$ & F=0.73 $\pm$ 0.03 \\
        $11$    & $(|01\rangle +|10\rangle)_{14}/\sqrt{2}$ & F=0.71 $\pm$ 0.04 \\
    \hline
   \end{tabular}\captionsetup{justification=raggedright}
\caption{Results from measuring qubit 2 and qubit 3 of the star cluster onto $Z_2Y_3$, which corresponds to a C-Phase($\pi/2$)$\cdot$CNOT$\cdot$C-Phase($\pi/2$) gate on states $|+0\rangle$ and $|+1\rangle$ up to a $(H_1H_4)(Z_4)^{s_2}(X_4)^{s_2+s_3+1}$.  
The fidelities of the tomographic reconstructed state with respect of the ideal state are reported. }
 \end{table}

 \begin{table}[h!]
  \centering
   \begin{tabular}{ >{\centering}m{1.2cm} >{\centering} m{3.7cm} m{2.4cm}<{\centering}}
   \hline
    $\mathbf{s_2} \mathbf{s_3} $ &  \textbf{Ideal Output State} & \textbf{Fidelity} \\  
   \hline
  $00$    & $(|00\rangle +|11\rangle)_{14}/\sqrt{2}$ & F=0.73 $\pm$ 0.04 \\
      $01$    & $(|00\rangle +|11\rangle)_{14}/\sqrt{2}$ & F=0.75 $\pm$ 0.04 \\
      $10$    & $(|01\rangle +|10\rangle)_{14}/\sqrt{2}$ & F=0.73 $\pm$ 0.04 \\
        $11$    & $(|01\rangle +|10\rangle)_{14}/\sqrt{2}$ & F=0.72 $\pm$ 0.03 \\
    \hline
   \end{tabular}\captionsetup{justification=raggedright}
\caption{Results from measuring qubit 2 and qubit 3 of the star cluster onto $Y_2Y_3$, which corresponds to a CNOT$\cdot$(I$\otimes$H$_y$) gate on states $|++_i\rangle$ and $|+-_i\rangle$ up to a $(X_4)^{s_3+1}$.  
The fidelities of the tomographic reconstructed state with respect of the ideal state are reported.}
 \end{table}

\newpage
.
\newpage
\section{Analysis of Verification Test on Four-qubit Linear Cluster State}
We report below the single probabilities during of the (1,3)-test and (2,4)-test on the linear cluster state. 
\begin{figure}[h!]
\begin{center}
\includegraphics[width=0.6\textwidth]{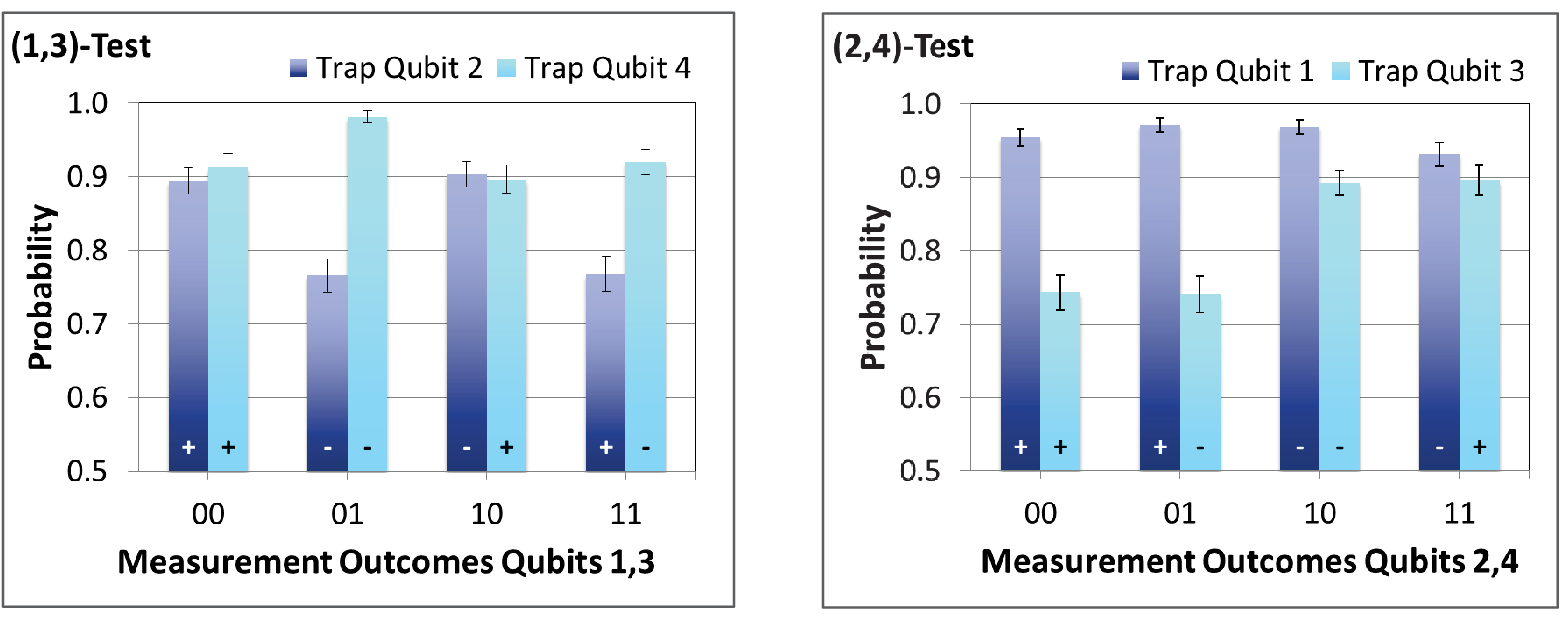}
\end{center} \captionsetup{justification=raggedright}
\caption{ Single probabilities of individual trap-qubits, corresponding to Alice's expected results, and according to respective measurement outcomes of the non-trap qubits (abscissa). At the bottom of each bar, the expected state of each trap-qubit is indicated. } 
\label{QST_histo}
\end{figure}
\section{Analysis of Verification Test on Four-qubit Star Cluster State}
\begin{figure}[h!]
\begin{center}
\includegraphics[width=0.6\textwidth]{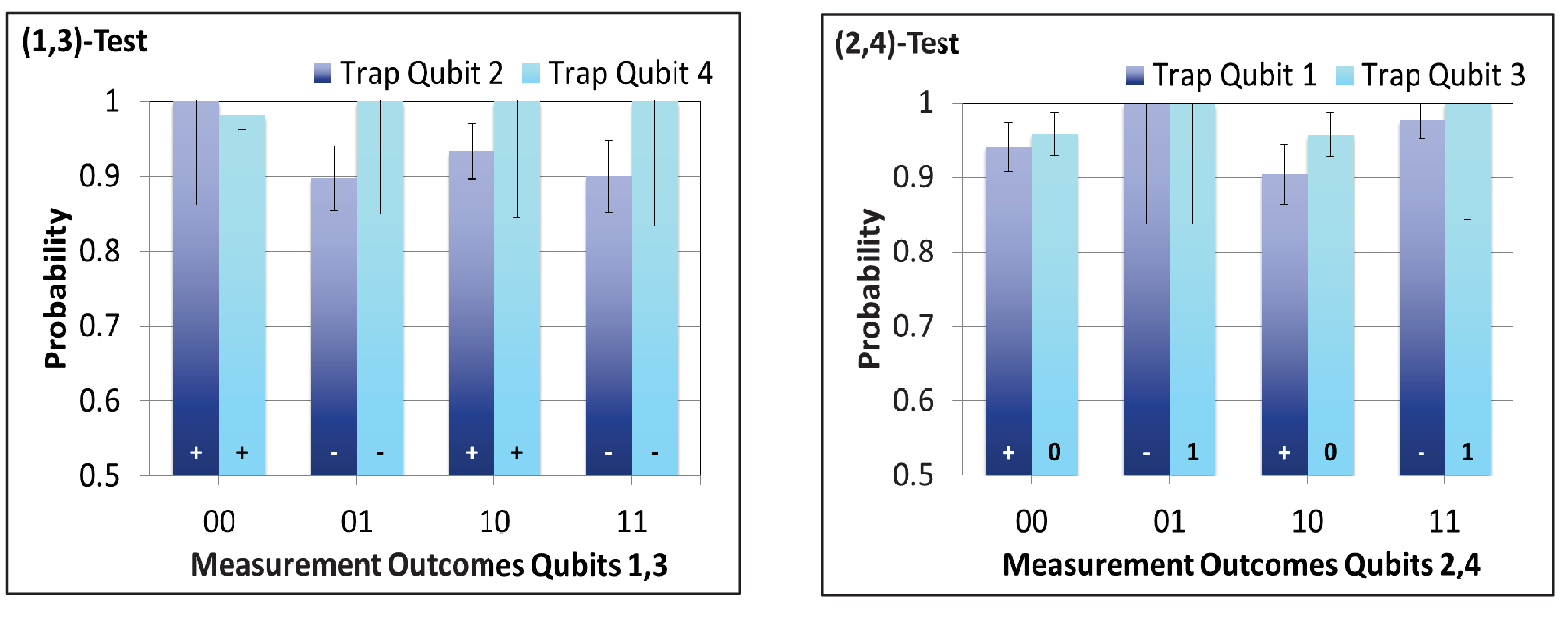}
\end{center} \captionsetup{justification=raggedright}
\caption{ Single probabilities per trap according to respective measurement outcomes. The error bars are increased here, due to the acquisition time of 600s per measurement.} 
\label{test1Star}
\end{figure}

The four-qubit star cluster state, used before for implementing entangling gates, is now used by Alice to verify Bob's dishonesty, even though the probability she's fooled is not 0, but exponentially small. \\
We report the two trap tests performed on the star cluster state, equivalently to the linear cluster case. 
 In this case in order to get two trap-qubits each time Alice measures  $Z_1Z_3$ (expecting trap qubits in $X$ basis) for (1,3)-test,   and $Z_2X_4$ (expecting one trap qubit in $X$ basis and one in $Z$) for (2,4)-test.  The single probabilities for the the two tests are shown in     Fig.\ref{test1Star}. 
